\documentclass[iop]{emulateapj}

\usepackage{times,natbib,graphicx,amsmath}

\shorttitle{Aquila X-1}
\shortauthors{Ludlam et al.}

\begin{document}

\title{Truncation of the Accretion Disk at One Third of the Eddington Limit in the Neutron Star Low-Mass X-ray Binary Aquila X-1}
\author{R. M. Ludlam\altaffilmark{1},
J. M. Miller\altaffilmark{1}, 
N. Degenaar\altaffilmark{2}, 
A. Sanna\altaffilmark{3},
E. M. Cackett\altaffilmark{4},
D. Altamirano\altaffilmark{5},
A. L. King\altaffilmark{6}
}
\altaffiltext{1}{Department of Astronomy, University of Michigan, 1085 South University Ave, Ann Arbor, MI 48109-1107, USA}
\altaffiltext{2}{Anton Pannekoek Institute for Astronomy, University of Amsterdam, Pastbus 94249, 1090 GE Amsterdam, The Netherlands}
\altaffiltext{3}{Dipartimento di Fisica, Universit\'{a} degli Studi di Cagliari, SP Monserrato-Sestu km 0.7, 09042 Monserrato, Italy}
\altaffiltext{4}{Department of Physics \& Astronomy, Wayne State University, 666 W. Hancock St., Detroit, MI 48201, USA}
\altaffiltext{5}{Department of Physics \& Astronomy, University of Southampton, Highfield, Southampton SO17 1BJ, UK}
\altaffiltext{6}{KIPAC, Stanford University, 452 Lomita Mall, Stanford, CA 94305 USA}

\begin{abstract} 
We perform a reflection study on a new observation of the neutron star low-mass X-ray binary Aquila X-1 taken with $\emph{NuSTAR}$ during the August 2016 outburst and compare with the July 2014 outburst. The source was captured at $\sim32\%\ L_{\mathrm{Edd}}$, which is over four times more luminous than the previous observation during the 2014 outburst.  Both observations exhibit a broadened Fe line profile. Through reflection modeling, we determine that the inner disk is truncated $R_{in,\ 2016}=11_{-1}^{+2}\ R_{g}$ (where $R_{g}=GM/c^{2}$) and $R_{in,\ 2014}=14\pm2\ R_{g}$ (errors quoted at the 90\% confidence level). Fiducial neutron star parameters (M$_{NS}=1.4$ M$_{\odot}$, $R_{NS}=10$ km) give a stellar radius of $R_{NS}=4.85\ R_{g}$; our measurements rule out a disk extending to that radius at more than the $6\sigma$ level of confidence. We are able to place an upper limit on the magnetic field strength of $B\leq3.0-4.5\times10^{9}$ G at the magnetic poles, assuming that the disk is truncated at the magnetospheric radius in each case. This is consistent with previous estimates of the magnetic field strength for Aquila X-1. However, if the magnetosphere is not responsible for truncating the disk prior to the neutron star surface, we estimate a boundary layer with a maximum extent of $R_{BL,\ 2016}\sim10\ R_{g}$ and $R_{BL,\ 2014}\sim6\ R_{g}$. Additionally, we compare the magnetic field strength inferred from the Fe line profile of Aquila X-1 and other neutron star low-mass X-ray binaries to known accreting millisecond X-ray pulsars. 
\end{abstract}


\section{Introduction}
Aquila X-1 is a neutron star (NS) residing in a low-mass X-ray binary (LMXB) that has exhibited X-ray pulsations, if intermittently. A LMXB consists of an accreting compact object with a companion star of approximately solar mass. The companion star in Aquila X-1 is categorized as a K0 V spectral type (\citealt{thorstensen78}; \citealt{sanchez17}). Coherent millisecond X-ray pulsations were detected for 150 s during persistent emission imply a spin frequency of 550 Hz \citep{casella08}. Type-I X-ray bursts place an upper limit on the distance to Aquila X-1 of 5.9 kpc away, assuming the bursts are Eddington limited \citep{jonker04}. 

The inclination of the system is constrained to be $<31^{\circ}$ by infrared photometry measurements performed by \citet{garcia99}. Intermittent dipping episodes may indicate an inclination as high as $72-79^{\circ}$ \citep{galloway16}. However, intermittent dipping may not be indicative of a high inclination. Another low inclination system, 4U 1543-47, exhibited intermittent dipping that was suggestive of an accretion instability \citep{park04}. Additionally, recent near-infrared spectroscopy rules out a high inclination and implies an inclination $23^{\circ}<i<53^{\circ}$ when considering conservative constraints \citep{sanchez17}. The magnetic field strength is estimated to be $0.4-31\times10^{8}$ G. This is inferred from pulsations signifying magnetically channeled accretion in $\emph{Rossi X-ray Timing Explorer}$ ($\emph{RXTE}$) observations \citep{mukherjee15}. Additionally, the \lq \lq propeller" phase, where material is thrown off from the disk at low luminosity and can no longer accrete onto the NS, implies a similar magnetic field strength (\citealt{campana98}; \citealt{asai13}).

Broadened and skewed Fe line profiles have been detected from accretion disks in NS LMXBs for the last decade (e.g. \citealt{BS07}; \citealt{papitto08}; \citealt{cackett08}, \citeyear{cackett10}; \citealt{disalvo09}; \citealt{Egron13}; \citealt{miller13}). These profiles are shaped from Doppler and relativistic effects \citep{Fabian89} and, as a consequence, the red wing can be used to determine the location of the inner edge of the disk. 

The accretion disk must extend down to or truncate prior to the surface of the NS. Disk truncation can occur above $\sim1\%$ L$_{\mathrm{Edd}}$ in one of two ways: either pressure balance between the accreting material and magnetosphere or a boundary layer of material extending from the surface. Below $\sim1\%$ L$_{\mathrm{Edd}}$, accretion in LMXBs can become inefficient and disk truncation can occur through other mechanisms, such as disk evaporation (\citealt{narayanyi95}; \citealt{tomsick09}; \citealt{degenaar17}). By studying sources with truncated accretion disks at sufficiently high L$_{\mathrm{Edd}}$, we can obtain estimates of magnetic field strengths (\citealt{IP09}; \citealt{cackett09}; \citealt{Pap09}; \citealt{miller11}; \citealt{degenaar14}, \citeyear{degenaar16}; \citealt{king16}; \citealt{ludlam16}) and/or extent of potential boundary layers (\citealt{PS01}; \citealt{king16}; \citealt{ludlam16}, \citealt{chiang16b}). 

It remains unclear whether the magnetic field is dynamically important in Aquila X-1 and other non-pulsating NS LMXBs. Aquila X-1 is frequently active with outbursts occurring about once a year (\citealt{campana13}; \citealt{waterhouse16}) making it a key target. \citet{king16} obtained observations of Aquila X-1 in the soft state with $\emph{NuSTAR}$ and $\emph{Swift}$ during the July 2014 outburst. They found that the disk was truncated at $15\pm3 \ R_{g}$ (where $R_{g}=GM/c^{2}$) at $\sim 7\%$ of the empirical Eddington luminosity ($L_{\mathrm{Edd}}=3.8\times10^{38}$ ergs s$^{-1}$; \citealt{kuulkers03}). This placed a limit on the strength of the equatorial magnetic field of $B<7\times10^{8}$ G that is consistent with previous estimates. 

The $\emph{Swift}$/BAT detected renewed activity on 2016 July 29 \citep{atel9287} that was confirmed to be a new outburst with a 500 s follow up $\emph{Swift}$/XRT observation \citep{atel9292}. Observations were taken with $\emph{NuSTAR}$ \citep{nustar} on 2016 August 7 when Aql X-1 was in the soft state at $\sim0.32\ L_{\mathrm{Edd}}$ during the outburst. We perform a reflection study on the prominent Fe K$_{\alpha}$ feature for this observation and compare with the 2014 outburst. 

\section{Observations and Data Reduction}
$\emph{NuSTAR}$ observations were taken of Aquila X-1 on 2014 July 17 and 18 (Obsids 80001034002 and 80001034003) and 2016 August 7 (Obsid 90202033002). Figure 1 shows the $\emph{Swift}$/BAT and MAXI daily monitoring lightcurves with vertical dashed lines to indicate when the $\emph{NuSTAR}$ observations were taken. Using the {\sc{nuproducts}} tool from {\sc nustardas} v1.5.1 with {\sc caldb} 20170503, we created  lightcurves and spectra for the 2016 observations. We used a circular extraction region with a radius of 100$^{\prime \prime}$ centered around the source and another region away from the source for the purpose of background subtraction. No Type-I X-ray bursts occurred during the 2016 observation. Initial modeling of the spectra with a constant fixed to 1 for the FPMA, found the floating constant for the FPMB to be within 0.95-1.05. We combine the two source spectra, background spectra, ancillary response matrices and redistribution matrix files via {\sc addascaspec} and {\sc addrmf}. Each of these have been weighted by exposure time. The 2014 observations were reduced using the most recent {\sc caldb}, 20170503, which has been updated since the reduction and analysis reported in \citet{king16}. The combined spectra were grouped  to have a minimum of 25 counts per bin \citep{cash} using {\sc grppha}. The net count rate for the combined spectra were 126.8 counts s$^{-1}$ in 2014 and 424.3 counts s$^{-1}$ in 2016.

\begin{figure}
\centering
\includegraphics[width=9.2cm]{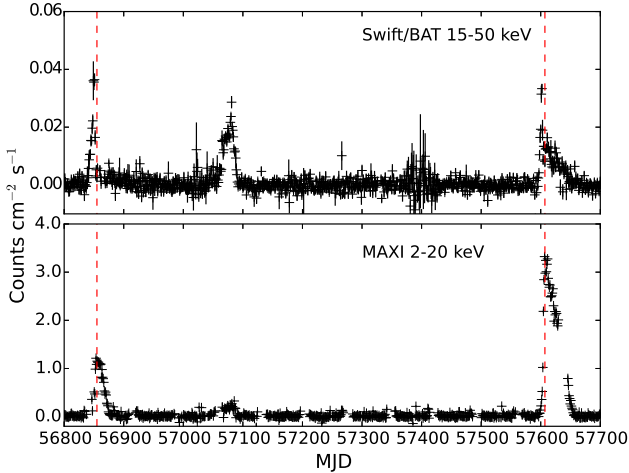}
\caption{Swift/BAT $15-50$ keV and MAXI $2-20$ keV daily monitoring lightcurves. The dashed lines represent the $\emph{NuSTAR}$ observations taken in July 2014 and August 2016. 
}
\label{fig:lc}
\end{figure}

We do not utilize the 2014 $\emph{Swift}$ observations as per \citet{king16} due to major flux differences between the $\emph{NuSTAR}$ and $\emph{Swift}$ spectra. The $\emph{Swift}$ spectrum required a multiplicative constant of 3.75 to match the $\emph{NuSTAR}$ flux.  This flux difference is likely due to the need to exclude the PSF core to avoid pile-up in the $\emph{Swift}$ data. Additionally, excluding the core of the PSF further limits the sensitivity of the $\emph{Swift}$ spectrum and, as a result, the reflection spectrum cannot be detected in the data. Furthermore, $\emph{Swift}$ only performed a short exposure observation (under 200 s) on the same day as the $\emph{NuSTAR}$ observation in 2016 that do not provide constraints. As a consequence, we opted to focus on the comparison of $\emph{NuSTAR}$ observations only in this study. 

\section{Spectral Analysis and Results}
We utilize XSPEC version 12.9.1 \citep{arnaud96} in this work with fits performed over the 3.0-30.0 keV energy range (the spectrum is dominated by background above 30 keV).  All errors were calculated using a Monte Carlo Markov Chain (MCMC) of length 100,000 and are quoted at 90$\%$ confidence level. We use {\sc tbnewer}\footnote{Wilms, Juett, Schulz, Nowak, in prep, http://pulsar.sternwarte.uni-erlangen.de/wilms/research/tbabs/index.html} to account for the absorption along the line of sight. Since $\emph{NuSTAR}$ has a limited lower energy bandpass it is unable to constrain the equivalent neutral hydrogen column density its own. We therefore set the equivalent neutral hydrogen column density to the \citet{dl90} value of $4.0\times10^{21}$ cm$^{-2}$. Moreover, this value is very close to column densities found with low energy spectral fitting to $\emph{XMM-Newton}$ and $\emph{Chandra}$ data \citep{campana14}.

\citet{king16} modeled the 2014 data using a Comptonized thermal continuum with  a relativistically blurred emergent reflection emission. 
We chose to forego this combination of models in an effort to provide a self-consistent approach between components. The reflection model in \citet{king16} assumes that a blackbody continuum is illuminating the disk, though the continuum is modeled with Comptonization. Further, the assumed blackbody in the reflection model that is providing the emergent reflection spectrum does not peak at the same energy as the Comptonized continuum. This means that the component  assumed to illuminate the accretion disk is not consistent with the emergent reflection spectrum. We chose to adopt a continuum model akin to \citet{lin07} for NS transients in the soft state. The continuum is described by two thermal components: a single temperature blackbody component ({\sc bbodyrad}) and a multi-temperature blackbody ({\sc diskbb}). The single temperature blackbody component is used to model the emission from the corona or boundary layer. The multi-temperature blackbody is used to account for the thermal emission from different radii in the accretion disk. The addition of a power-law component may be needed in some cases and is suggestive of weak Comptonization.

Initial fits were performed with two thermal components, which gave a poor fit in each case ($\chi^{2}_{2014}/d.o.f.=4088.70/591$ and $\chi^{2}_{2016}/d.o.f.=3946.47/585$), partly due to the presence of strong reflection within the spectrum. We added a power-law component with the photon index  bound at a hard limit of 4.0. Steep indices of this nature have been observed in \citet{sobczak00} and \citet{park04} for black hole X-ray binaries. The additional power-law component  improved the the overall fit at more than the $9\sigma$ level of confidence, as determined via F-test, in each case. However, the reflection is still unaccounted for by this model. The broadened Fe K emission line can be seen in Figure 2 for each outburst. 

\begin{figure}
\centering
\includegraphics[width=9.2cm]{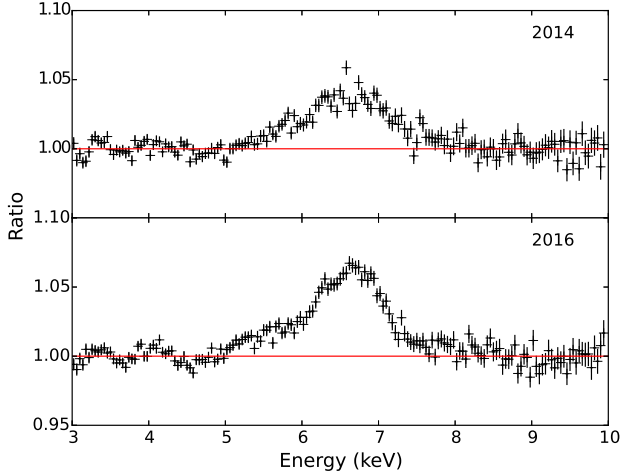}
\caption{Comparison of Fe line profiles for Aql X-1 during the 2014 and 2016 outbursts created by taking the ratio of the data to the continuum model. The continuum model was fit over the energies of $3.0-5.0$ keV and $8.0-10.0$ keV. The iron line region was ignored ($5.0-8.0$ keV) to prevent the feature from skewing the fit. Ignoring above 10.0 keV gives an unhindered view of the Fe K$_{\alpha}$ line, though it models both the continuum and some reflection continuum.
}
\label{fig:feline}
\end{figure}

We account for the emergent reflection from an ionized disk by convolving {\sc reflionx}\footnote{http://www-xray.ast.cam.ac.uk/$\sim$mlparker/reflionx$\_$models/reflionx$\_$bb.mod} \citep{reflionx} with the relativistic blurring kernal {\sc relconv} \citep{relconv}. The {\sc reflionx} model has been modified to assume the disk is illuminated by a blackbody. We tie the blackbody temperature of the reflection and continuum emission.  We use a constant emissivity index, $q$, fixed at 3 as would be expected for an accretion disk illuminated by a point source in an assumed geometry of flat, Euclidean space (\citealt{wilkins12}). Different  geometries, such as a boundary layer surrounding the NS or hot spots on the surface illuminating the disk, replicate the same $r^{-3}$ emissivity profile (D. Wilkins, priv. comm.). The iron abundance, $A_{Fe}$, is fixed at half solar abundance in agreement with the previous analysis on Aql X-1 \citep{king16}. We fix the dimensionless spin parameter, $a_{*}$ (where $a_{*}=cJ/GM^{2}$), to 0.259 which is implied from the pulsation spin frequency of 550 Hz  (\citealt{braje00}; \citealt{casella08}; \citealt{king16}). This assumes a NS mass of 1.4 M$_{\odot}$, radius of 10 km, and a moderately soft equation of state \citep{braje00}. The inner disk radius, $R_{in}$, is given in units of innermost stable circular orbit (ISCO). We convert this value to $R_{g}$ given that 1 ISCO $= 5.2\ R_{g}$ for $a_{*}=0.259$ \citep{bardeen72}.  

The  {\sc xspec} model we used for each spectrum was {\sc tbnewer}*({\sc diskbb}+{\sc bbodyrad}+{\sc pow}+{\sc relconv}*{\sc reflionx}). This provided an improvement in the overall fit at more than the 25$\sigma$ level of confidence ($\chi^{2}_{2014}/d.o.f.=620.29/583$ and $\chi^{2}_{2016}/d.o.f.=603.08/579$) over the prior model that did not account for reflection within the spectra. Figure 3 shows the best fit spectra and model components. Model parameters and values are listed in Table 1. The exact nature of the power-law component is unknown as it may or may not be physical, but it is statistically needed at more than the $15\sigma$ level of confidence for each case.

\begin{table}
\caption{Aql X-1 Reflionx Modeling}
\label{tab:aqlrefl} 
\begin{center}
\begin{tabular}{llcc}
\hline
Component & Parameter &2014&2016 \\
\hline
{\sc tbnewer}
&$N_\mathit{H} (10^{22}) ^{\dagger}$
&$0.4$
&$0.4$
\\
{\sc diskbb}
&$kT$
&$1.64\pm0.02$
&$1.69_{-0.02}^{+0.01}$
\\
&norm 
&$12.0_{-0.5}^{+0.3}$
&$62\pm2$
\\
{\sc bbodyrad}
&$kT$
&$2.27\pm0.02$
&$2.33_{-0.02}^{+0.01}$
\\
&norm
&$1.2\pm0.1$
&$4.1_{-0.2}^{+0.4}$
\\
{\sc powerlaw}
&$\Gamma$
&$3.7\pm0.1$
&$3.96_{-0.21}^{+0.03}$
\\
&norm
&$1.2\pm0.1$
&$4.8_{-0.9}^{+0.2}$
\\
{\sc relconv}
&$q ^{\dagger}$
&3.0
&3.0
\\
&$a_{*} ^{\dagger}$
&0.259
&0.259
\\
&$\mathit{i} (^{\circ})$
&$26_{-3}^{+2}$
&$26\pm2$
\\
&$R_\mathit{in} (ISCO) $
&$2.7\pm0.4$
&$2.1_{-0.2}^{+0.3}$
\\
&$R_\mathit{in} (R_\mathit{g}) $
&$14\pm2$
&$11_{-1}^{+2}$
\\
&$R_\mathit{out} (R_\mathit{g}) ^{\dagger}$ 
&400
&400
\\
{\sc reflionx}
&$\xi$
&$400_{-40}^{+60}$
&$200\pm10$
\\
&$A_\mathit{Fe} ^{\dagger}$
&$0.5$
&$0.5$
\\
&$\mathit{z} ^{\dagger}$
&0
&0
\\
&norm
&$0.25_{-0.03}^{+0.02}$
&$3.5\pm0.2$
\\
&$F_{unabs,\ 0.5-50.0\ keV}$
&$6\pm1$
&$29_{-6}^{+4}$
\\
&$L_{ 0.5-50.0\ keV}$
&$2.5\pm0.4$
&$12_{-3}^{+2}$
\\
&$L_{ 0.5-50.0\ keV}/L_{\mathrm{Edd}}$
&$0.07\pm0.01$
&$0.32_{-0.08}^{+0.05}$
\\
\hline
&$\chi_\nu^{2}$(d.o.f.)
&1.06 (583)  
&1.04 (579)
\\
\hline
$^{\dagger}$ = fixed
\end{tabular}

\medskip
Note.--- Errors are quoted at $90\%$ confidence level. The $N_{H}$ was fixed to the \citet{dl90} value for the absorption column density along the line of sight and given in units of cm$^{-2}$. The {\sc reflionx} model used has been modified to for an accretion disk illuminated a blackbody. The blackbody temperatures were tied between the continuum and reflection emission. The power-law index was pegged at a hard limit of 4.0. Flux is given in units of $10^{-9}$ ergs cm$^{-2}$ s$^{-1}$. Luminosity is calculated at a maximum distance of 5.9 kpc and given in units of $10^{37}$ ergs s$^{-1}$. $L_{\mathrm{Edd}}=3.8\times10^{38}$ ergs s$^{-1}$ \citep{kuulkers03}. For reference, 1 ISCO $= 5.2\ R_{g}$ for $a_{*}=0.259$.
\end{center}
\end{table}

\begin{figure}
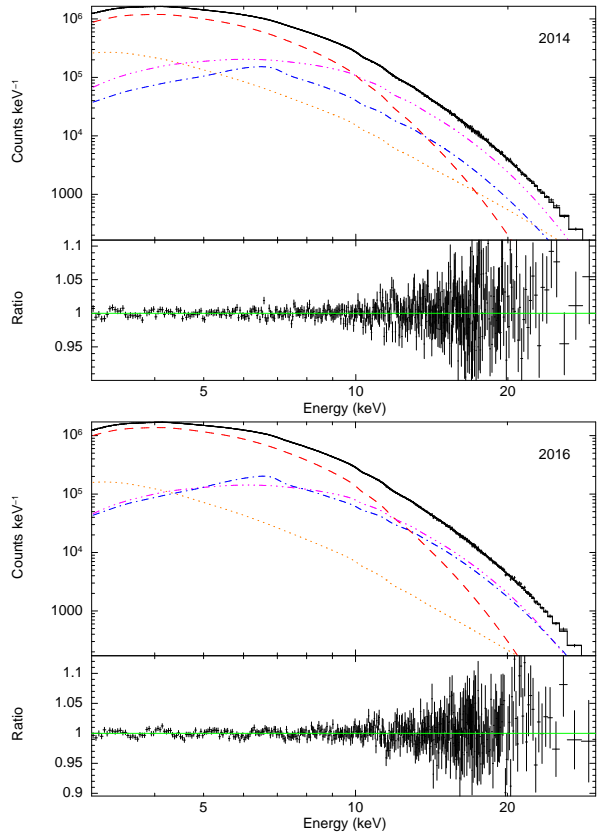

\centering
\includegraphics[angle=270,width=8.2cm]{2014comps_rebin.eps}
\includegraphics[angle=270,width=8.2cm]{2016comps_rebin.eps}
\caption{Aql X-1 spectrum fit from 3.0-30.0 keV with a {\sc diskbb} (red dash line), {\sc blackbody} (purple dot dot dot dash line), power-law (orange dot line), and {\sc reflionx} (blue dot dash line).  The ratio of the data to the model is shown in the lower panel. The data were rebinned for clarity. Table 1 lists parameter values for each model.
}
\label{fig:comps}
\end{figure}

For the data taken during the 2014 outburst, the {\sc diskbb} component has a temperature of $kT=1.64\pm0.02$ keV and norm$=12.0_{-0.5}^{+0.3}$ km$^{2}$/100\ kpc$^{2}$ cos($i$).  The {\sc bbodyrad} component has a temperature of $kT=2.27\pm0.02$ keV and normalization of $1.2\pm0.1$ km$^{2}$/100\ kpc$^{2}$. The power-law has a steep photon index of $\Gamma=3.7\pm0.1$ with a normalization of $1.2\pm0.1$ photons keV$^{-1}$ cm$^{-2}$ s$^{-1}$ at 1 keV.  The inner disk radius is truncated at $R_{in}=2.7\pm0.4$ ISCO ($14\pm2\ R_{g}$). The inclination was found to be $26_{-3}^{+2}\ ^{\circ}$. 

For the data taken during the 2016 outburst, the {\sc diskbb} component has a temperature of $kT=1.69_{-0.02}^{+0.01}$ keV and norm$=62\pm2$ km$^{2}$/100\ kpc$^{2}$ cos($i$).  The {\sc bbodyrad} component has a temperature of $kT=2.33_{-0.02}^{+0.01}$ keV and normalization of $4.1_{-0.2}^{+0.4}$ km$^{2}$/100\ kpc$^{2}$. Again, the photon index is steep at $\Gamma=3.96_{-0.21}^{+0.03}$ with a normalization of $4.8_{-0.9}^{+0.2}$  photons keV$^{-1}$ cm$^{-2}$ s$^{-1}$ at 1 keV. The inner disk radius is truncated at $R_{in}=2.1_{-0.2}^{+0.3}$ ISCO ($11_{-1}^{+2}\ R_{g}$). The inclination is $26\pm2\ ^{\circ}$, which also agrees with the previous observation. 

The blackbody and disk blackbody normalizations in both fits are implausibly small when used to infer a radial extent of the emitting region. This systematic underestimation was proposed by \citet{london86} to be the result of spectral hardening as photons travel through an atmosphere above pure blackbody emission and is supported through numerical simulations (\citealt{shimura95}; \citealt{merloni00}). The consistency in model parameter values with only the normalization changing between the two soft state observations likely indicates similar accretion geometries. We allow the emissivity parameter to be free to check if our results are dependent on the emissivity index being fixed at 3. The emissivity index tends towards a slightly higher value of $q=3.1$ for the 2014 observation and $q=2.5$, which is consistent with the disk extending down to a smaller radii in the most recent observation. All model parameters are consistent within the $3\sigma$ level of confidence with those reported in Table 1.
Figure 4 shows how the goodness-of-fit changes with inner disk radius for each observation. We use the XSPEC \lq \lq steppar" command to determine how the goodness-of-fit changed as a function of inner disk radius. At each evenly placed step, $R_{in}$ was fixed while the other parameters were free to adjust to find the best fit. The ISCO is ruled out at more than the $6\sigma$ level of confidence in each case.

\begin{figure}
\centering
\includegraphics[width=9.2cm]{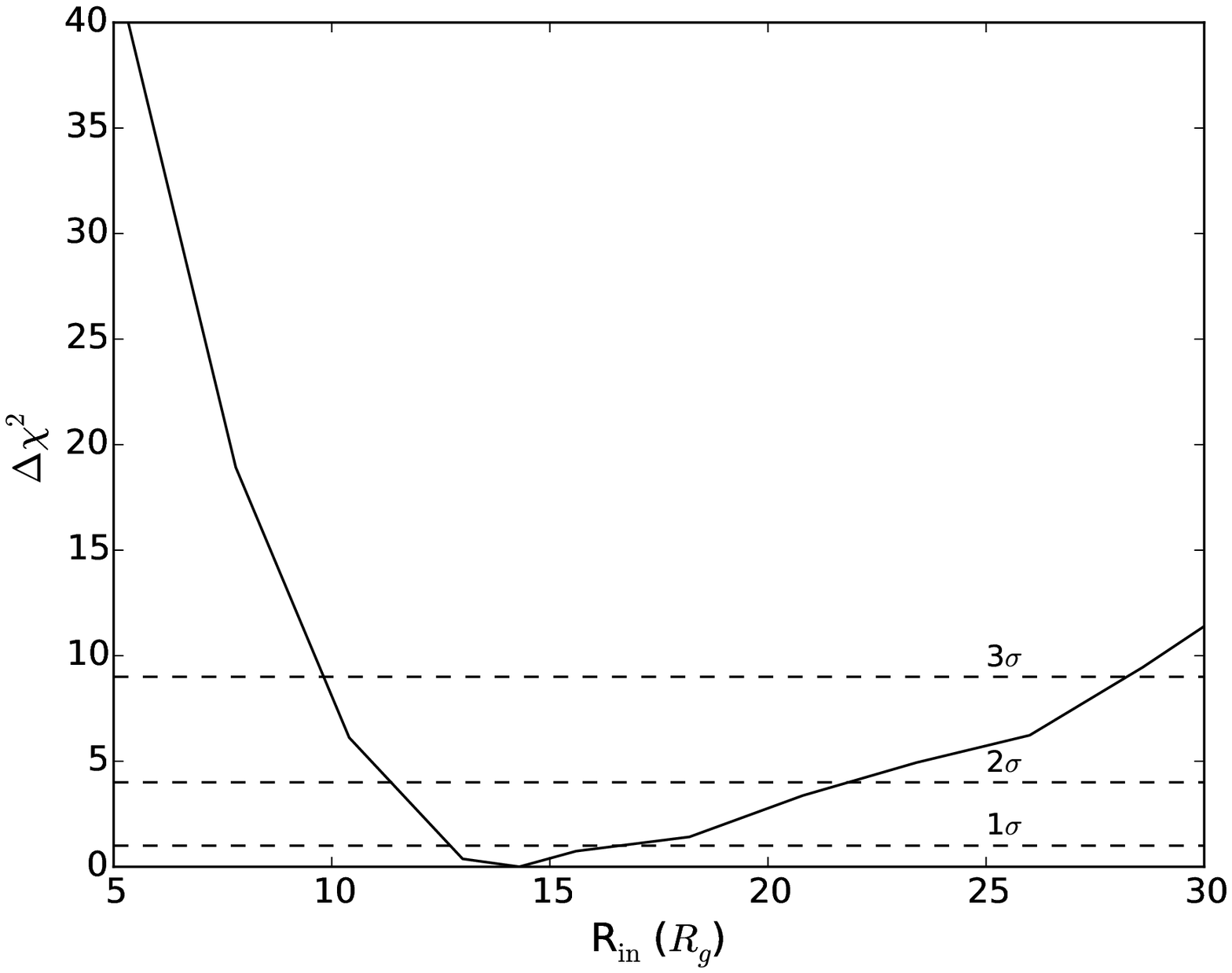}
\includegraphics[width=9.2cm]{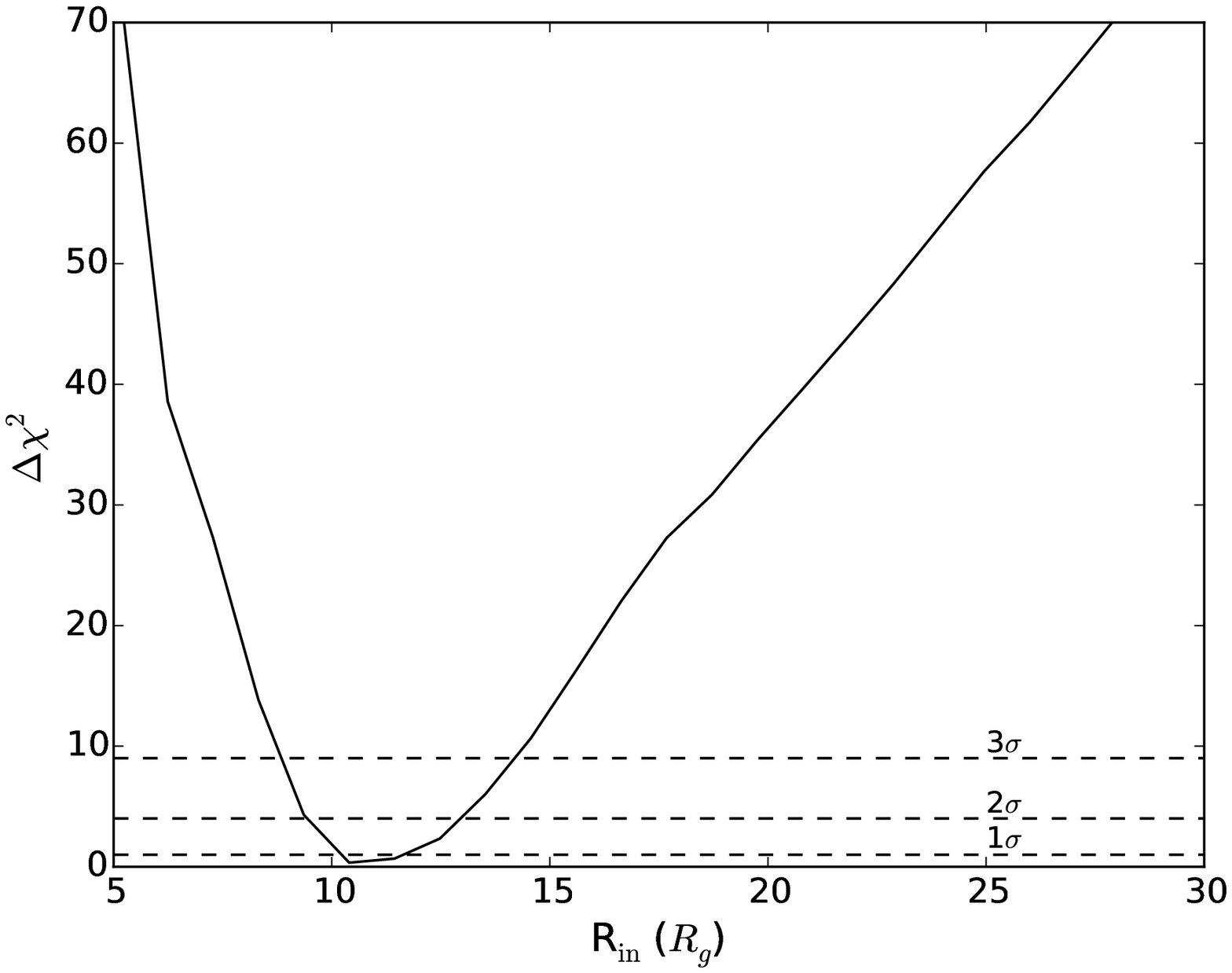}
\caption{Change in goodness-of-fit with inner disk radius for the 2014 (top) and 2016 (bottom) outbursts taken over evenly spaced steps generated with XSPEC \lq \lq steppar". The inner disk radius was held constant while the other parameters were free to adjust to find the minimum $\chi^2$ value at each step. The dashed lines represent the $1\sigma$, $2\sigma$, and $3\sigma$ confidence intervals.
}
\label{fig:feline}
\end{figure}

\section{Discussion}
We present a new observation of Aquila X-1 taken with $\emph{NuSTAR}$ during its August 2016 outburst and compare it to the July 2014 outburst. We perform reflection fits that indicate the disk is truncated prior to the surface of the neutron star. The location of the inner disk radius during the 2014 observation is $14\pm2\ R_{g}$. This is consistent with the previous results found in \citet{king16}, although we modeled the continuum in a different way.
The location of the inner disk radius remains truncated ($11_{-1}^{+2}\ R_{g}$) during the 2016 observation even though the flux is over four times larger.
Additionally, both spectra imply an inclination of $26\pm2^{\circ}$ which is consistent with infrared photometric and spectroscopic measurements (\citealt{garcia99}; \citealt{sanchez17}).

By assuming that the ram pressure in the disk is balanced by the outward pressure of the magnetic field, we can place an upper limit on the magnetic field strength using the maximum extent the inner disk of $R_{in}=13\ R_{g}$ from the 2016 spectrum. Assuming a mass of 1.4 M$_{\odot}$, taking the maximum distance to be 5.9 kpc, and using the maximum unabsorbed flux from $0.5-50.0$ keV of $33\times10^{-9}$ erg cm$^{-2}$ s$^{-1}$ as the bolometric flux,  the magnetic dipole moment, $\mu$, can be estimated from Equation (1): 
\begin{equation}
\small
\begin{aligned}
\mu = 3.5 \times 10^{23} \ k_{A}^{-7/4} \ x^{7/4} \left(\frac{M}{1.4\ M_{\odot}}\right)^{2} \\ \times \left(\frac{f_{ang}}{\eta} \frac{F_{bol}}{10^{-9}\ \mathrm{erg \ cm^{-2} \ s^{-1}}}\right)^{1/2} \frac{D}{3.5\ \mathrm{kpc}} \ \mathrm{G\ cm}^{3}
\end{aligned}
\end{equation}
with $x$ being the number of gravitational radii (\citealt{IP09}; \citealt{cackett09}). If we assume an accretion efficiency of $\eta=0.2$ and unity for the angular anisotropy, $f_{ang}$, and conversion factor, $k_{A}$, then $\mu\sim6.7\times10^{26}$ G cm$^{3}$. For a NS of 10 km, this implies a magnetic field strength at the poles of $B\leq1.3\times10^{9}$ G.  Alternatively, if we assume a different conversion factor between disk and spherical accretion of $k_{A}=0.5$ as proposed in \citet{long05}, the strength of the magnetic field increases to $B\leq4.5\times10^{9}$ G.  For the 2014 outburst, we use the upper limit of $R_{in}=16\ R_{g}$ and the maximum unabsorbed flux from $0.5-50.0$ keV of $7\times10^{-9}$ erg cm$^{-2}$ s$^{-1}$ to place a limit on the magnetic field strength to be $B\leq0.9\times10^{9}$ G for $k_{A}=1.0$ and $B\leq3.0\times10^{9}$ for $k_{A}=0.5$. Note that the magnetic field strength at the equator is half as strong as at the pole. \citet{king16} found a similar value for the maximum strength of the magnetic field for Aquila X-1 of $B\simeq1.4\times10^{9}$ G at the magnetic poles.  We report the upper limit on the magnetic field strength using the conversion factor of $k_A=0.5$ hereafter since it encompasses the value for $k_A=1.0$.

If, however, the magnetosphere was not responsible for truncating the disk, a boundary layer extending from the surface of the NS could plausibly halt the accretion flow. Equation 2, taken from \citet{PS01}, provides a way to estimate the maximum radial extent of this region from the mass accretion rate. 
\begin{equation}
\small
\begin{aligned}
\log \left(R_{max}-R_{NS}\right) \simeq 5.02+0.245\left| \log \left(\frac{\dot{M}}{10^{-9.85}\ \mathrm{M}_{\odot}\ \mathrm{yr}^{-1}}\right) \right| ^{2.19}
\end{aligned}
\end{equation}
We determine the mass accretion rate using the unabsorbed luminosity from $0.5-50.0$ keV and an accretion efficiency of $\eta=0.2$ to be $1.1_{-0.3}^{+0.1}\times10^{-8}\ \mathrm{M}_{\odot}\ \mathrm{yr}^{-1}$ during the 2016 observation and $2.2\pm0.4\times10^{-9}\ \mathrm{M}_{\odot}\ \mathrm{yr}^{-1}$ during the 2014 observation. This gives a maximum radial extent of $\sim10\ R_{g}$ for the boundary layer during 2016 and $\sim6\ R_{g}$ during 2014 (assuming canonical values of $\mathrm{M}_{NS}=1.4\ \mathrm{M}_{\odot}$ and $R_{NS}=10$ km). This is consistent with the location of the inner disk radius during the 2016 outburst, but falls short of the inner disk radius in our 2014 fits. \citet{king16} found a similar radial extent of the boundary layer of $\sim7.8\ R_{g}$, but this can be increased by rotation of the NS or a change in viscosity to be consistent with the truncation radius.

It is more likely that the magnetic field is responsible for disk truncation in this source. The equatorial magnetic field strength inferred from the Fe line profile ($B\leq15.0-22.5\times10^{8}$ G) is consistent with other estimates of the magnetic field strength ($0.4-31\times10^{8}$ G: \citealt{campana98}; \citealt{asai13}; \citealt{mukherjee15}) and are well within the range to truncate an accretion disk (\citealt{mukherjee15}). Following Equation (1) and rearranging for inner disk radius in terms of flux, the inner disk radius should scale like $F_{bol}^{-2/7}$. Thus for magnetic truncation the inner disk radius should decrease as the flux increases, which is what we see for the different observations. Conversely, if the boundary layer were responsible for disk truncation in each case, we should see the inner disk radius increase. Additionally, the maximum extent of the boundary layer during the 2014 observation does not agree with the location of the inner disk radius, pointing to the magnetic field being a more probable explanation for disk truncation. Moreover, although the extent of the boundary layer is consistent with the inner disk radius in the 2016 fits, the behavior of decreasing inner disk radius with increasing flux is indicative of magnetic truncation.

\subsection{Comparison of Magnetic Field Strengths}
$\emph{NuSTAR}$ has observed a number of NS LMXBs with Fe lines that imply truncated disks. This has provided a means of placing an upper limit on the strength of their magnetic fields, assuming the disk is truncated at the Alfv\'{e}n radius (where the ram pressure of the accreting material is balanced by the magnetic pressure outwards). The implied magnetic field strengths reside between $10^{8}-10^{9}$ G and are similar to accreting millisecond X-ray pulsars (AMXPs). \citet{mukherjee15} systematically estimated the upper and lower limits to the equatorial magnetic field strengths of 14 known AMXPs using $\emph{Rossi X-ray Timing Explorer}$ ($\emph{RXTE}$). They used the highest flux that the source exhibited pulsations and the radius of the NS to determine $B_{min}$ and the lowest flux that exhibited pulsations and corotation radius with the disk to determine $B_{max}$ in each case.

\begin{figure}[t]
\centering
\includegraphics[width=8.2cm]{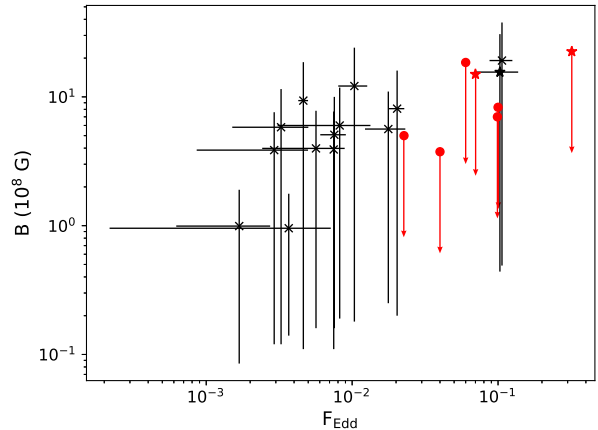}
\caption{Comparison of equatorial magnetic field strengths of NSs in LMXBs (red) inferred from Fe line profiles to known AMXPs (black) reported in \citet{mukherjee15} versus Eddington fraction. The stars represent estimates for Aquila X-1. See Table 2 for magnetic field strengths and Eddington fraction values.}
\label{fig:mag}
\end{figure}

\begin{table}
\caption{Magnetic Field Strengths Versus Eddington Fraction}
\label{tab:BvsEdd}
\begin{center}
\begin{tabular}{llcc}
\hline
Source & $B$ ($10^{8}$ G)& $F_{Edd}$&ref.\\
\hline
Swift J1756.9-2508 & $0.18-24.1$& $0.008-0.013$& 1\\
XTE J0929-314 & $0.12-11.5$&$0.002-0.005$ & 1 \\
XTE J1807.4-294 &$0.11-18.6$ & $0.004-0.005$& 1\\
NGC 6440 X-2 & $0.12-7.6$& $0.001-0.005$&1\\
IGR J17511-3057&$0.19-11.8$ &$0.003-0.013$ & 1\\
XTE J1814-338 & $0.16-7.8$&$0.002-0.009$ & 1\\
HETE J1900.1-2455 & $0.16-10.0$ & $0.006-0.009$ & 1\\
SAX J1808.4-3658 & $0.14-1.77$&$0.0002-0.0071$ & 1\\
IGR J17498-2921 & $0.20-16.0$& $0.018-0.023$& 1\\
XTE J1751-305 & $0.25-11.0$& $0.012-0.023$& 1\\
SAX J1748.9-2021 & $0.49-37.8$& $0.09-0.12$& 1\\
Swift J1749.4-2807 & $0.11-7.7$ & $0.0074-0.0075$ & 1\\
IGR J00291+5934 & $0.085-1.9$& $0.001-0.003$ & 1\\
Aquila X-1 & $0.44-30.7$ & $0.07-0.13$& 1\\
&$\leq15.0$&$0.07$&\\
&$\leq22.5$&$0.32$&\\
1RXS J180408.9-3 &$\leq5.0$&$0.02$& 2\\
&$\leq8.3$&$0.10$& 3\\
4U 1705-44&$\leq7.0$& $0.10$& 4\\
XTE J1709-267 & $\leq3.8-18.5$ & $0.04-0.06$ & 5 \\
\hline
\end{tabular}

\medskip
Note.--- (1) \citealt{mukherjee15}; (2) \citealt{ludlam16}; (3) \citealt{degenaar16}; (4) \citealt{ludlam17a}; (5) \citealt{ludlam17b}.  All Eddington fraction values are a lower limit since we used the maximum Eddington luminosity value of $L_{\mathrm{Edd}}=3.8\times10^{38}$ ergs s$^{-1}$ \citep{kuulkers03}.
\end{center}
\end{table}

Figure 5 presents a comparison of magnetic field strengths of known AMXPs to NS LMXBs observed with $\emph{NuSTAR}$ versus Eddington fraction, $F_{\mathrm{Edd}}$. As can be seen, the NS LMXBs populate higher values of Eddington fraction. Each point from \citet{mukherjee15} represents a range in magnetic field strength and $F_{\mathrm{Edd}}$ that the AMXP lies and does not embody an actual measurement. Values can be found in Table 2. The advantage of magnetic field strengths inferred from the Fe line profiles using $\emph{NuSTAR}$ is that they do not suffer from pile-up or instrumental effects until a source reaches $\sim10^{5}$ counts s$^{-1}$. We use the maximum Eddington luminosity of $3.8\times10^{38}$ ergs s$^{-1}$ from \citet{kuulkers03} when calculating the Eddington fraction for each source. If the Eddington luminosity is smaller, all points would be shifted to higher values of Eddington fraction. Therefore, these are all lower limits. 

Another caveat of this comparison is that pulsations have not been detected yet for the sources observed with $\emph{NuSTAR}$. For Aquila X-1 in particular the 2014 observation is within the same $F_{\mathrm{Edd}}$ range as the observation taken by $\emph{RXTE}$ when pulsations were detected. Additionally, our upper limit on the strength of the magnetic field agrees with the estimate when pulsations were detected. It is clear that the strengths implied from Fe line profiles are valuable and consistent with those seen for AMXPs. Therefore, Fe lines can be used to estimate magnetic field strengths to first order.

\section{Summary}
We present a reflection study of Aquila X-1 observed with $\emph{NuSTAR}$ during the July 2014 and August 2016 outbursts. We find the disk to be truncated prior to the surface of the NS at $14\pm2\ R_{g}$ during 2014 observation when the source was at 7\% of Eddington and $11_{-1}^{+2}\ R_{g}$ during the 2016 observation when the source was at 32\% of Eddington. This implies an upper limit on the strength of the magnetic field at the poles of $3.0-4.5\times10^{9}$ G, if the magnetosphere is responsible for truncating the disk in each case. If a boundary layer is responsible for halting the accretion flow instead, we estimate the maximal radial extent to be $\sim6\ R_{g}$ for the 2014 observation and $\sim10\ R_{g}$ during 2016. These values can be increased through viscous and spin effects, but the behavior of decreasing inner disk radius with increasing flux favors magnetic truncation. Finally, when comparing the strength of magnetic fields in NS LMXBs to those of known AMXPs we find that they are consistent while probing a higher value of Eddington fraction.
\\ \\ \\
We thank the referee for their prompt and thoughtful comments that have improved the quality of this work. This research has made use of the NuSTAR Data Analysis Software (NuSTARDAS) jointly developed by the ASI Science Data Center (ASDC, Italy) and the California Institute of Technology (Caltech, USA). ND is supported by a Vidi grant from the Netherlands Organization for Scientific Research (NWO). EMC gratefully acknowledges support from the National Science Foundation through CAREER award number AST-1351222. DA acknowledges support from the Royal Society.


\begin{thebibliography}{}
\bibitem[Arnaud(1996)]{arnaud96}Arnaud, K. A.\ 1996, in Astronomical Society of the Pacific Conference Series, Vol. 101, Astronomical Data Analysis Software and Systems V, ed.\ G. H. Jacoby \& J. Barnes, 17
\bibitem[Asai et al.(2013)]{asai13}Asai, K., Matsuoka, M., Mihara, T., et al.\ 2013, ApJ, 773, 117
\bibitem[Bardeen et al.(1972)]{bardeen72}Bardeen, J. M., Press, W. H., \& Teukolsky, S. A.\ 1972, ApJ, 178, 347
\bibitem[Bhattacharyya \& Strohmayer(2007)]{BS07}Bhattacharyya, S., \& Strohmayer, T. E. \ 2007 \apj, 664, L103
\bibitem[Braje et al.(2000)]{braje00}Braje, T. M., Romani, R. W., \& Rauch, K. P.\ 2000, ApJ, 531, 447
\bibitem[Cackett et al.(2008)]{cackett08}Cackett, E. M., Miller, J. M., Bhattacharyya, S., Grindlay, J. E., Homan, J., van der Klis, M., Miller, M. C., Strohmayer, T. E., \& Wijnands, R.\ 2008, ApJ, 674, 415
\bibitem[Cackett et al.(2009)]{cackett09}Cackett, E. M., Altamirano, D., Patruno, A., Miller, J. M., Reynolds, M., Linares, M., \& Wijnands, R.\ 2009, \apj, 694, L21
\bibitem[Cackett et al.(2010)]{cackett10}Cackett, E. M., Miller, J. M., Ballantyne, D. R., Barret, D., Bhattacharyya, S., Boutelier, M., Coleman Miller, M. Strohmayer, T. E., \& Wijnands, R.\ 2010, ApJ, 720, 205
\bibitem[Campana et al.(1998)]{campana98}Campana, S., Stella, L., Mereghetti, S., et al.\ 1998, ApJ, 499, 65
\bibitem[Campana et al.(2013)]{campana13}Campana, S., Coti Zelati, F., \& D'Avanzo, P.\ 2013, MNRAS, 432, 1695
\bibitem[Campana et al.(2014)]{campana14}Campana, S., Brivio, F., Degenaar, N., et al. \ 2014, MNRAS, 441, 1984
\bibitem[Casella et al.(2008)]{casella08}Casella, P., Altamirano, D., Patruno, A., Wijnands, R., \& van der Klis, M.\ 2008, ApJ, 674, 41
\bibitem[Cash(1979)]{cash}Cash, W. \ 1979, ApJ, 228, 939
\bibitem[Chiang et al.(2016)]{chiang16b}Chiang, C.-Y., Morgan, R. A., Cackett, E. M., et al.\ 2016, ApJ, 831, 45
\bibitem[Dauser et al.(2010)]{relconv}Dauser, T., Wilms, J., Reynolds, C. S., \& Brenneman, L. W. \ 2010, MNRAS, 409, 1534
\bibitem[Degenaar et al.(2016)]{degenaar16}Degenaar, N., Altamirano, D., Parker, M., et al.\ 2016, MNRAS, 461, 4049
\bibitem[Degenaar et al.(2017)]{degenaar17}Degenaar, N., Pinto, C., Miller, J. M., et al.\ 2017, MNRAS, 464, 398
\bibitem[Degenaar et al.(2014)]{degenaar14}Degenaar, N., Miller, J. M., Harrison, F. A., Kennea, J. A., Kouveliotou, C., \& Younes, G. \ 2014, ApJ, 796, L9
\bibitem[Di Salvo et al.(2009)]{disalvo09}Di Salvo, T. et al. \ 2009, MNRAS, 398, 2022
\bibitem[Dickey \& Lockman(1990)]{dl90}Dickey, J. M., \& Lockman, F. J.\ 1990, ARA\&A, 28, 215
\bibitem[Egron et al.(2013)]{Egron13}Egron, E. et al. 2013, A\&A, 550, A5
\bibitem[Fabian et al.(1989)]{Fabian89}Fabian, A. C., Rees, M. J., Stella, L., \& White, N. E.\ 1989, MNRAS, 238, 729
\bibitem[Galloway et al.(2016)]{galloway16}Galloway, D. K., Ajamyan, A. N., Upjohn, J., \& Stuart, M.\ 2016, MNRAS, 461, 3847
\bibitem[Garcia et al.(1999)]{garcia99}Garcia, M. R., Callanan, P. J., McCarthy, J., Eriksen, K., \& Hjellming, R. M.\ 1999, ApJ, 518, 422
\bibitem[Harrison et al.(2013)]{nustar}Harrison, F. A., Craig, W. W., Christensen, F. E., et al.\ 2013, ApJ, 770, 103
\bibitem[Ibragimov \& Poutanen(2009)]{IP09}Ibragimov, A., \& Poutanen, J. 2009, MNRAS, 400, 492
\bibitem[Jonker \& Nelemans(2004)]{jonker04}Jonker, P. G., \& Nelemans, G.\ 2004, MNRAS, 354, 355
\bibitem[King et al.(2016)]{king16}King, A. L. et al. \ 2016, ApJ, 819, L29
\bibitem[Kuulkers et al.(2003)]{kuulkers03}Kuulkers, E., den Hartog, P. R., in't Zand, J. J. M., Verbunt, F. W. M., Harris, W. E., \& Cocchi, M. \ 2003, A\&A, 399, 663
\bibitem[Lin et al.(2007)]{lin07}Lin, D., Remillard, R. A., \& Homan, J.\ 2007, ApJ, 667, 1073
\bibitem[London et al.(1986)]{london86}London, R. A., Taam, R. E., Howard, W. M.\ 1986, ApJ, 306, 170L
\bibitem[Long et al.(2005)]{long05}Long, M., Romanova, M. M., \& Lovelace, R. V. E.\ 2005, ApJ, 634, 1214
\bibitem[Ludlam et al.(2016)]{ludlam16}Ludlam, R. M., Miller, J. M., Cackett, E. M., et al. 2016, ApJ, 824, 37
\bibitem[Ludlam et al.(2017a)]{ludlam17a}Ludlam, R. M., Miller, J. M., Bachetti, M., et al.\ 2017a, ApJ, 836, 140
\bibitem[Ludlam et al.(2017b)]{ludlam17b}Ludlam, R. M., Miller, J. M., Cackett, E. M., Degenaar, N., Bostrom, A. C.\ 2017b, ApJ, 838, 79
\bibitem[Mata S\'{a}nchez et al.(2017)]{sanchez17}Mata S\'{a}nchez, D., Mu\~{n}oz-Darias, T., Casares, J., \& Jim\'{e}nez-Ibarra, F.\ 2017, MNRAS, 464, 41
\bibitem[Merloni et al.(2000)]{merloni00}Merloni, A., Fabian, A. C., Ross, R. R.\ 2000, MNRAS, 313, 193
\bibitem[Miller et al.(2011)]{miller11}Miller, J. M., Maitra, D., Cackett, E. M., Bhattacharyya, S., \& Strohmayer, T. E. \ 2011, ApJ, 731, L7
\bibitem[Miller et al.(2013)]{miller13}Miller, J. M. et al. \ 2013, ApJ, 779, L2
\bibitem[Miller et al.(1998)]{Miller98}Miller, M. C., Lamb, F. K., \& Psaltis, D. \ 1998, ApJ, 508, 791
\bibitem[Mukherjee et al.(2015)]{mukherjee15}Mukherjee, D., Bult, P., van der Klis, M., \& Bhattacharya, D.\ 2015, MNRAS, 452, 3994
\bibitem[Narayan \& Yi(1995)]{narayanyi95}Narayan R., Yi I., 1995, ApJ, 452, 710
\bibitem[Papitto et al.(2008)]{papitto08}Papitto, A., D'Ai, A., Di Salvo, T., Iaria, R., Menna, M. T., Burderi, L., \& Riggio, A.\ 2008, The Astronomer's Telegram, 1846, 1
\bibitem[Papitto et al.(2009)]{Pap09}Papitto, A., Di Salvo, T., D'Ai, A., Iaria, R., Burderi, L., Riggio, A., Menna, M. T., \& Robba, N. R. \ 2009, A\&A, 493, L39
\bibitem[Park et al.(2004)]{park04}Park, S. Q., Miller, J. M., McClintock, J. E., Remillard, R. A., Orosz, J. A., et al.\ 2004, ApJ, 610, 378
\bibitem[Popham \& Sunyaev(2001)]{PS01}Popham, R., \& Sunyaev, R. 2001, ApJ, 547, 355
\bibitem[Ross \& Fabian(2005)]{reflionx}Ross, R. R., \& Fabian, A. C. \ 2005, MNRAS, 358, 211
\bibitem[Sanna et al.(2016a)]{atel9287}Sanna, A., Riggio, A., Pintore, F.\ et al. \ 2016a, The Astronomer's Telegram, 9287, 1
\bibitem[Sanna et al.(2016b)]{atel9292}Sanna, A., Riggio, A., Pintore, F.\ et al. \ 2016b, The Astronomer's Telegram, 9292, 1
\bibitem[Shimura \& Takahara(1995)]{shimura95}Shimura, T., Takahara, F.\ 1995, ApJ, 445, 780
\bibitem[Sobczak et al.(2000)]{sobczak00}Sobczak, G. J., McClintock, J. E., Remillard, R. A., Cui, W., Levine, A. M, et al.\ 2000, ApJ, 544, 993
\bibitem[Thorstensen et al.(1978)]{thorstensen78}Thorstensen, J., Charles, P., \& Bowyer, S.\ 1978, ApJL, 220, L131
\bibitem[Tomsick et al.(2009)]{tomsick09}Tomsick J., Yamaoka K., Corbel S., Kaaret P., Kalemci E., Migliari S., 2009, ApJL, 707, L87
\bibitem[Waterhouse et al.(2016)]{waterhouse16}Waterhouse, A. C., Degenaar, N., Wijnands, R., et al.\ 2016, MNRAS, 456, 4001 
\bibitem[Wilkins \& Fabian(2012)]{wilkins12}Wilkins, D. R., Fabian, A. C., 2012, MNRAS, 424, 1284
\bibitem[Zhang et al.(1998)]{zhang98}Zhang, W., Jahoda, K., Kelley, R. L., Strohmayer, T. E., Swank,
J. H., \& Zhang, S. N.\ 1998, ApJL, 495, L9
\end{thebibliography}
\end{document}